# Comparative Performance of Machine Learning Algorithms in Cyberbullying Detection: Using Turkish Language Preprocessing Techniques

Emre Cihan ATES[a*], Erkan BOSTANCI[b], Mehmet Serdar GÜZEL[c]

[a] Department of Security Science, Gendarmerie and Coast Guard Academy (JSGA), Ankara, Turkey.
[b] Department of Computer Engineering, Ankara University, Ankara, Turkey.
[c] Department of Computer Engineering, Ankara University, Ankara, Turkey.

## Abstract

With the increasing use of the internet and social media, it is obvious that cyberbullying has become a major problem. The most basic way for protection against the dangerous consequences of cyberbullying is to actively detect and control the contents containing cyberbullying. When we look at today's internet and social media statistics, it is impossible to detect cyberbullying contents only by human power. Effective cyberbullying detection methods are necessary in order to make social media a safe communication space. Current research efforts focus on using machine learning for detecting and eliminating cyberbullying. Although most of the studies have been conducted on English texts for the detection of cyberbullying, there are few studies in Turkish. Limited methods and algorithms were also used in studies conducted on the Turkish language. In addition, the scope and performance of the algorithms used to classify the texts containing cyberbullying is different, and this reveals the importance of using an appropriate algorithm. The aim of this study is to compare the performance of different machine learning algorithms in detecting Turkish messages containing cyberbullying. In this study, nineteen different classification algorithms were used to identify texts containing cyberbullying using Turkish natural language processing techniques. Precision, recall, accuracy and F1 score values were used to evaluate the performance of classifiers. It was determined that the Light Gradient Boosting Model (LGBM) algorithm showed the best performance with 90.788% accuracy and 90.949% F1 Score value.

***Keywords — Cyberbullying, machine learning, social media, crime, natural language processing***

## 1. Introduction

With the rapid development of social networks today, the scope of internet use has significantly expanded. We are in a period in which people create a virtual personality on social media for purposes such as expressing their emotions and sharing things with other people (Mulyadi and Fitriana, 2018).

[*] Corresponding author.
E-mail addresses: emre_cihan_ates@hotmail.com(E.C.Ates), ebostanci@ankara.edu.tr(E.Bostanci), mguzel@ankara.edu.tr(M.S.Guzel).

After the virtual personality formation, which usually starts with membership to social media sites, data types such as text, sound, image and video can be transferred. In addition to data transfer, social media is a communication environment where people can also send messages such as messages, status or comments to each other.

The rapid increase in the use of social media over the past years has brought many positive and negative developments (Akdemir and Lawless, 2020; Siddiqui and Singh, 2016). One of the most common negative effects is cyberbullying. Cyberbullying can be defined as the act of intentionally violating another person through technological communication tools (Englander et al., 2017; Peter and Petermann, 2018). Harassment or aggression through e-mails, mobile phones, online games and social media can be considered cyberbullying.

Since, most studies focusing on detecting cyberbullying are English-based, the Turkish language based studies are quite insufficient. Turkish language, which is being used in Turkey first and foremost, and also many parts of the world, is a head-final language, and that is one of the main factors that make it difficult to do researches with it (Çakir and Güldamlasioğlu, 2016; Çöltekin, 2020).

For example, particularly in Turkey where Turkish is spoken as a mother tongue, when the rate of social media platforms usage in 2020 on which cyberbullying is made are examined (WeAreSocial, 2020);

- We see that it's the 6th country on Twitter with its 11.800.000 users,
- the 5th country on Instagram with its 38,000,000 users,
- the 10th country on Facebook with its 37,000,000 users,
- and that it's the 15th country on social media platforms with an average usage time of 2 hours 51 minutes a day.

When social media usage statistics in Turkey in particular are examined, it reveals how widespread is the use of social media is in Turkey. Therefore, it has become obligatory to detect cyberbullying in Turkish texts. In this study, it is aimed to compare the performance of different machine learning algorithms in cyberbullying detection by using natural language processing techniques specific to Turkish language. Following the introduction of the study, a background on which the concepts of cyber bullying, machine learning, natural language processing and Turkish language are defined respectively has been prepared. In the related works section, studies in the literature on cyberbullying detection are presented. In the methodology section, information was given about the data used for detecting cyberbullying, pre-processing and machine learning models used. In the result and discussion section, the results obtained after machine learning modelling are shared and compared with other studies in the literature. In the conclusion and future works section, the gains obtained after the study and future work plans to increase cyberbullying detection performance in Turkish texts are presented.

## 2. Background

### 2.1. Cyberbullying

Cyberbullying is the state of aggressive and repetitive negative behaviours using technological devices against another person or group with an imbalance of power (Brailovskaia et al., 2018; Englander et al., 2017). The harm caused by cyberbullying is more serious than physical bullying. Items subject to bullying can be dispersed very quickly on the internet and social media environment. Moreover, parallel to the rapid development of technology, the ability to hide the criminals' identities by different methods in virtual platforms and therefore the difficulty in finding criminals creates a situation that motivates criminals to do such acts.

Today, regarding the processing of cyberbullying on social media; some cyberbullying behaviours include sending threatening and vulgar messages, disclosing private personal information, creating online gossip, sending repetitive swearing and insulting messages and humiliating any person or group. Cyberbullying can occur with all genders and at different age groups, and the act of cyberbullying can be related to physical, cultural, racial and even religious prejudices (Ferreira et al., 2016; Muralidharan and La Ferle, 2018).

It is known that many people experience physical and mental victimization after cyberbullying behaviour. These behaviours also gradually reduce people's tendency to trust social media (Marzano, 2019). For this reason, detecting cyberbullying on social media is a must more than a need. When the Twitter sample of 2020 is examined, an average of 194,444 tweets were sent in different languages every 1 minute (Lewis and Callahan, 2020). In a 24-hour period sample, the average number of tweets sent reached 279,999,360, and it is almost impossible to physically examine such a large amount of data in different languages. For this reason, today's studies focus primarily on normalization of text with natural language processing techniques and subsequently classification based on machine learning.

### 2.2. Machine Learning

Machine learning is the learning ability of a computer to make decisions on its own, based on the existing data and experiences (Hutter et al., 2019). Decisions to be taken within the scope of machine learning are generally for prediction or classification purposes (Sarkar, 2019). Within the scope of this study, machine learning was used to check whether Twitter posts contain cyberbullying or not. In the context, available data is known as training data. The computer trains the models and classifies test data using learning algorithms. There are different learning methods according to the labelling of the training data (Aggarwal, 2018). What we mean by labelling is the classification of data by human power before machine learning. If the learning model is dependent on tagged data, then this model is supervised learning. Unsupervised learning occurs when training data are not tagged. In unsupervised learning, the machine learns to classify itself according to the similarities and differences between the data. The learning to be applied on the combination of tagged and untagged data is semi-supervised learning.

Machine learning algorithms are widely used to detect messages containing cyberbullying in social networks (Rosa et al., 2019). Its usage is in the form of supervised learning in general and in this study. There are different learning methods according to the labelling of the training data, and the algorithms used in this study are explained below.

- **Naive Bayes:** There is an assumption in this algorithm that the inputs are independent from each other and of equal importance. Although it seems quite simple in terms of the way it works, successful and fast results are obtained in many examples adapted to real life (Liu et al., 2017; Rogel-Salazar, 2018). The fact that the input data is independent means that each attribute is affected by a certain number of inputs and has no significance among them. This situation is usually almost impossible. In general, there are three types: Gaussian Naive Bayes, Multinomial Naive Bayes and Bernoulli Naive Bayes.
- **Gaussian Naïve Bayes:** If the features in the data are continuous value, it is assumed that these values are sampled from a Gaussian distribution (normal distribution) (Yang, 2019). This method is quite easy for estimating the distribution of data.
- **Multinomial Naive Bayes:** It is used for classifying data into multi-class category (considering the frequency of words) (Ott, 2013).
- **Bernoulli Naive Bayes:** Tough similar to multinomial Naive Bayes, it only classifies data into (Boolean) two categories (Yang, 2019).
- **Decision Tree Model:** It is a widely used method in prediction and descriptive models due to its easy interpretation, understanding and compatibility with databases (Alpaydin, 2010; Binkhonain and Zhao, 2019). As the name suggests, it is a popular estimation method that look like a tree and can classify it separately for each branch (Ozdemir, 2016). In this model, the data set is divided into smaller groups and a decision tree is formed.
- **Extra Trees Classifier:** In this model, a series of decision trees are created from various subsets of the training data set (Rogel-Salazar, 2018). It is a classification in which averages are used to check the accuracy of the prediction by estimating the model that fits the random decision tree.
- **Linear Discriminant Analysis (LDA):** This algorithm is used as a dimension reduction technique in the preprocessing step for classification and machine learning applications (Siqueira et al., 2017). The goal is to avoid over compliance and also to reduce computational costs.
- **Quadratic Discriminant Analysis (QDA):** The way this algorithm works is similar to the Linear Discriminant Analysis algorithm, but the difference is that QDA assumes that each class has its covariance matrix (Siqueira et al., 2017).
- **AdaBoost (Adaptive Boosting):** The basic rule of this algorithm is to fit a number of students into the frequently changing versions of the data (Mohri et al., 2018). Then the estimations, voting method was used in order to get estimates from all students. The data is re-weighted and the learning algorithm is reapplied for each iteration. The purpose here is to transform weak classifiers into a strong classifier.

- **Gradient Boosting Model:** In this algorithm, as in the AdaBoost algorithm, weak learning is expected to turn into strong learning by learning from mistakes, and the main difference is that the same purpose is done through lost functions (Sarkar, 2019).

- **Random Forest:** In this algorithm, it is aimed to obtain a more robust and accurate modelling with multiple decision tree modelling (Rogel-Salazar, 2018; Sarkar, 2019). By reducing the variance in the way of the study, it is ensured that accuracy is increased as a community-based decision tree by preventing potential over-fitting.

- **Logistic Regression Model:** This algorithm is a machine learning algorithm by examining the relationship between dependent variables and independent variables, just like in linear regression, and the biggest difference from classical linear regression is data classification (Rogel-Salazar, 2018; Sarkar, 2019). It is a special form of regression, and although the name of the algorithm is referred to as regression, it is used for classification rather than estimation. For this reason, logistic regression method is also known as binary classification (Ozdemir, 2016). Logistic regression has a low variance due to its simple operation structure and is therefore less prone to over fitting.

- **Perceptron Neural Network:** This algorithm works in a single or multi-layer artificial neural network system and is actively used in supervised learning where classification processes are performed (Yang, 2019). Single layer sensors perform the classification process by giving the most appropriate weight coefficients to the inputs linearly.

- **Support Vector Machine (SVM):** This algorithm works especially on the basis of classification, it is the algorithm used due to its high generalization performance (Liu et al., 2017; Mohri et al., 2018). Although the learning time is long, more optimized results can be obtained due to the fact that it reduces the number of processes in learning, prevents memorization style learning more, and is successful on large-volume data sets.

- **Linear Support Vector Classification:** This algorithm is used for linearly separable data and gives fast results, especially in large data (Naicker et al., 2020; Yang, 2019). Classification is done using straight lines, which shows that the data can be separated linearly.

- **XGBoost:** As with classical gradient enhancement models, it is a decision tree-based community learning algorithm (Rufaida et al., 2020). It was revealed in 2016 and it's considered new. The main difference from other algorithms lies in its scalability, which enables fast learning through parallel and distributed computing and provides efficient memory usage. It is free from over fitting and bias.

- **K - Nearest Neighbours Model:** It is a machine learning algorithm used to classify data points by calculating the distances between different data points (Liu et al., 2017). The predicted class in the test data is the class to which most of the closest neighbours in the training set belong (Rogel-Salazar, 2018). Therefore, the purpose of the classification is to group in the closest position. When predictions are required and data diversity is low, the results can be quite successful.

- **Stochastic Gradient Descent:** It is an optimization algorithm that is used to find the parameter values (coefficients) of a function (f) and minimizes the cost function (Yang, 2019). Although the usage method is quite simple, it can give effective results. Since only one sample is randomly selected from the data set for each iteration in terms of the way it works, the way the algorithm takes to reach the minimum is usually noisier than your typical Gradient Descent algorithm.

- **Light Gradient Boosting Model:** This algorithm uses a tree-based learning algorithm and is one of the most powerful gradient boosting algorithms in machine learning (Rufaida et al., 2020). It performs better compared to XGBoost and CatBoost (Al Daoud, 2019). It was revealed in 2016 and it's considered new. The biggest difference with XGBoost is experienced during tree-based learning. LGBM does not grow line by line at tree-level-wise, but instead at leaf-wise level.

- **Voting Classifier Model**: This algorithm combines two or more machine learning algorithms and predicts the class or tag based on the best average probability rated from all classifiers (Kumar et al., 2017). It is one of the best performing algorithms because it combines the properties of different classifiers.

## 2.3. Natural Language Processing

Natural language processing (NLP) is the process of making unstructured natural language data detectable by the computer using various techniques and algorithms (Sarkar, 2019). The natural language we use consists of symbolic expressions, and the basic element that distinguishes these expressions from each other in the article is the characters we call "letters". The characters come together to form words and sentences. There is a direct relationship between characters and words. For example, the expression "sun" describes a totally different object, while the expression "Sunday", which contains the same characters, refers to a day in the calendar. For this reason, every word made up of characters can evoke a different subject in the human brain. There are a number of rules for using languages, such as subject + verb + object. However, people can informally form sentences and perceive each other without obeying these rules. The greatest challenge for natural language processing is that human language is highly ambiguous, and it is in constant change and development. The basic criterion to overcome this difficulty is to teach the symbolic words and characters of the language to the machines along with the basic rules of the language in question. The more successful the learning, the more successful will be for machines to perceive human language.

Natural language processing today; linguistics is seen as a common branch of artificial intelligence and computer science, and this situation positively affects the learning of language by machines (Lane et al., 2019). Every day, studies are carried out on the basis of Acoustic - Phonetic,

Morphological - Syntactic and Semantic - Pragmatic in order to better understand human language (Bengfort et al., 2018). Natural language processing, which was very limited especially in the 1940s, has been successfully used in many fields such as machine translation, text summarization, question answering, information extraction, subject modelling, idea mining, text correction, optical character recognition and text recognition emotion analysis (Lane et al., 2019).

**2.4. Turkish Language**

Turkish is a language spoken today in south east Europe and western Asia. According to Ethnologue (2020), Turkish is the 17th most spoken language in the world with approximately 85 million speakers, while it is the 9th most used language in web content according to WeAreSocial (2020). The Latin alphabet is used to characterize the letters in Turkish, but the letters Ç, Ş, Ğ, I, İ, Ö, Ü, which are not originated in the Latin alphabet and considered derived later, are used in Turkish, besides that the letters q, w and x are not used in the Turkish language. It is a written language, read and written from left to right, and its alphabet consists of 29 letters.

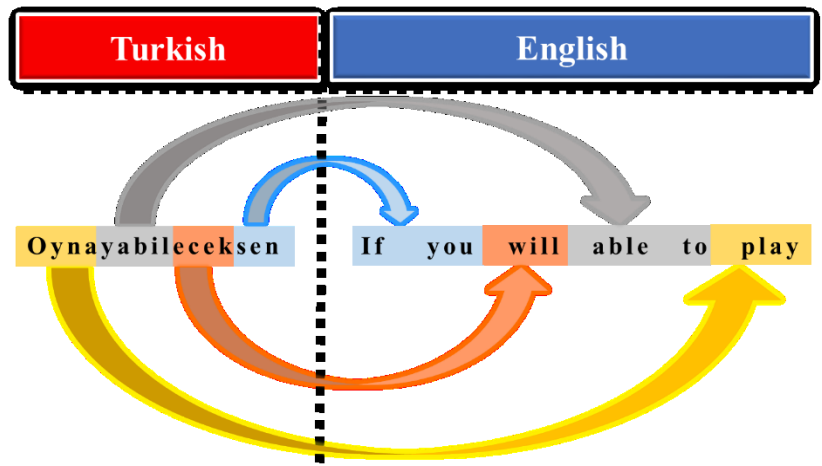

**Fig. 1.** Spelling of the same expression in Turkish and English

As can be seen in fig.1, Turkish is a suffixed language. A phrase that is expressed in a long form in English can be expressed in a shorter form with the suffixes at the end of the word in Turkish. This situation is related to the change in the integrity of meaning by adding suffixes to the end of the word (Dehkharghani et al., 2017). However, Turkish being a final head language caused the morphological structure of the language to be more complex. Due to the difficulty of reaching the root of the words and the complexity of the conjugations, there're not many studies conducted on machine learning in the Turkish language.

**3. Related Works**

In line with the purpose of detecting cyberbullying from Turkish social media posts, studies on Turkish natural language processing and cyberbullying detection were examined in the literature. Due to the insufficient formation of Turkish data sets, models in different languages, especially English, were included in the literature search.

- Raza et al. (2020) used Logistic Regression, Naïve Bayes, Random Forest, Adaboost and Voting classifier algorithms to detect cyberbullying messages in English. The data set used shows an uneven distribution in labelling. The best performance value has been obtained with 84.4% accuracy in the Voting algorithm.

- Van Hee et al. (2018) utilizing the n-gram method, only used the linear support vector machines algorithm, to detect cyberbullying messages in English and Danish. The dataset shows an uneven distribution, with 4.73% for English and 6.97% for Danish. Cyberbullying messages are also divided into groups (Curse, Defamation, Defence, Encouragement, Insult, Sexual, Threat), and the main classification is based on whether the messages contain cyberbullying or not. The best F1 Score was achieved with 64% for English and 61% for Danish.

- Husain (2020) tried to detect the use of offensive language on social media messages in Arabic. TF-IDF was used in the pre-processing of the data, and the emojis were also taken into account in the study. The data set used shows an uneven distribution. SVM, logistic regression, and decision tree, bagging, AdaBoost, and random forest algorithms were used within the scope of the study. Using the bagging algorithm, the best performance value was obtained with 88% F1 Score.

- Kanan et al. (2020) K - Nearest Neighbours, Support Vector Machines, Naïve Bayes, Random Forest, and decision trees J48 were used in the study to detect cyberbullying and cyber harassment messages in Arabic. The data set used consists of Facebook and Twitter data and shows an uneven distribution. The best performance value obtained using the random forest algorithm was 92.9% F1 score.

- By Liu et al. (2019) Naïve Bayes, Decision Tree, Random Forest, Tree Ensemble, Logistic Regression, Support Vector Machines algorithms were used in the study to detect cyberbullying in English texts. 9 different datasets, including social media such as Facebook, Twitter and Formspring were used. The best performance value was obtained with an accuracy of 81.75% using the logistic regression algorithm.

- In the study conducted by Muneer and Fati (2020) on the detection of cyberbullying, Logistic Regression, Light Gradient Boosting Machine, Stochastic Gradient Descent, Random Forest, AdaBoost, Naive Bayes, and Support Vector Machines algorithms are used. Using the logistic regression algorithm, it obtained the best performance value with 92.8% F1 score, and the LGBM algorithm obtained 92.7% F1 score.

- Talpur and O'Sullivan (2020) in the study of cyberbullying by Naïve Bayes, K-Nearest Neighbours, Decision Tree, Random Forest, and Support Vector Machine algorithms are used. Cyberbullying tweets are classified according to their risk levels, and SMOTE technique has been used against unbalanced class distribution. The best performance value was obtained with a 92.9% F1 score using the random forest algorithm.

- By Ali and Syed (2020) in the study on detecting cyber bullying in English texts Support Vector Machine, Naïve Bayes, Random Forest and ensemble approach were used. The best performance value was obtained with 79.3% accuracy using SVM algorithm.

- Bayesian Logistic Regression, Random Forest Algorithm, Multi-Layer Algorithm, J48 Algorithm and Support Vector Machine were used in the study conducted by Altay and Alatas (2018) to detect cyberbullying in English texts. The best performance value was 72.9% F1 score obtained using the Bayesian Logistic Regression algorithm.

- By Zhang et al. (2019) Linear Support Vector Machine, Logistic Regression, Decision Tree, Random Forest, Gradient Boosting, Perceptron, and Logistic Regression algorithms were used in the study conducted to detect cyberbullying in Japanese texts. The data were taken from Twitter and show a balanced distribution. The best performance value was obtained by using the Gradient Boosting algorithm was 93.5% F1 Score.

All of the above-mentioned studies are studies in languages other than Turkish. In these studies, various classification algorithms were used to detect cyberbullying, and pre-processing techniques unique to each language were used. The difficulty of natural language processing due to the fact that the Turkish language is a head final language is the main reason for the limited work in this field. In this context, when the limited number of studies in the literature on cyberbullying detection in Turkish language are examined;

- In the study by Çöltekin (2020) on the offensive use of Turkish language, social media messages, 19% of which were labelled as offensive language, were used as a data set. Machine learning was used to detect offensive language. By using only Linear Support Vector Machine classifiers algorithm, 77.3% F1 Score was achieved.

- In the study conducted by Bozyiğit et al. (2019) on Turkish cyberbullying detection, artificial neural networks were used. Existing libraries were not used in Turkish natural language processing. By creating a "Harmful Terms" list, using the Levenshtein algorithm between the term and the input, the misspellings were tried to be corrected. TF-IDF was used in text mining. In the study where the number of hidden layers were predicted as 2, 91% F1 Score was obtained.

- In the study regarding the detection of cyber bullying in Turkish conducted by Özel et al. (2017), Support Vector Machines, Decision Tree (C4.5), Naïve Bayes Multinomial, and K- Nearest Neighbours (kNN) algorithms were used. Existing libraries were not used due to their poor performance in natural language processing of Turkish. 84% accuracy was achieved with Multinomial Naïve Bayes.

As stated above, studies on Turkish cyberbullying detection are very limited. A limited number of studies have been conducted on the basis of machine learning algorithms in the studies. For this reason, there has been a need for studies to detect cyberbullying in Turkish texts on the basis of different algorithms.

## 4. Methodology

The aim of this study is to compare the performance of machine learning algorithms that can be used in the detection of Turkish content cyberbullying messages from social media platforms in the Twitter sample.

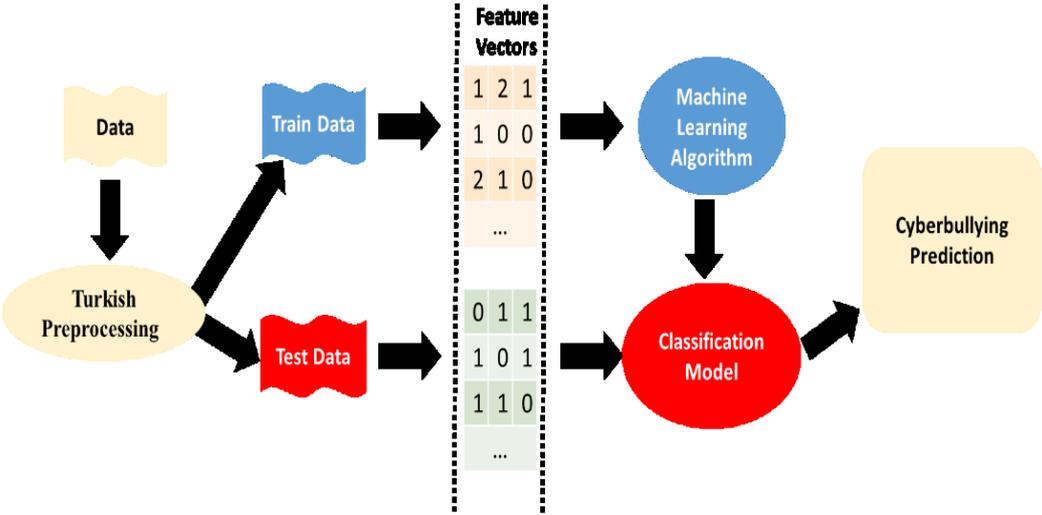

**Fig. 2.** Cyberbullying Detection Model

This study has been conducted over the cyberbullying detection model in fig 2, and the procedures regarding the methodology are explained in the subtitles.

### 4.1. Data Description

For the detection of cyberbullying, data set created on Twitter by Bozyiğit (2018) was used. There are 3000 twitter messages in the data set, 1500 of which contain cyberbullying (Cyberbullying = 1), and 1500 do not contain cyberbullying (Cyberbullying = 0). Equal number of labelling is important for a more balanced distribution in classification based on machine learning.

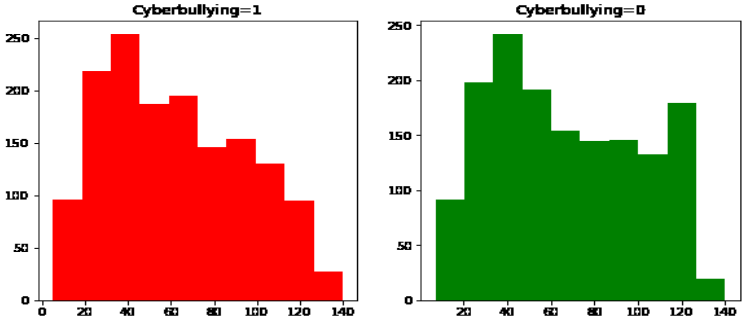

**Fig. 3.** Characters in tweets

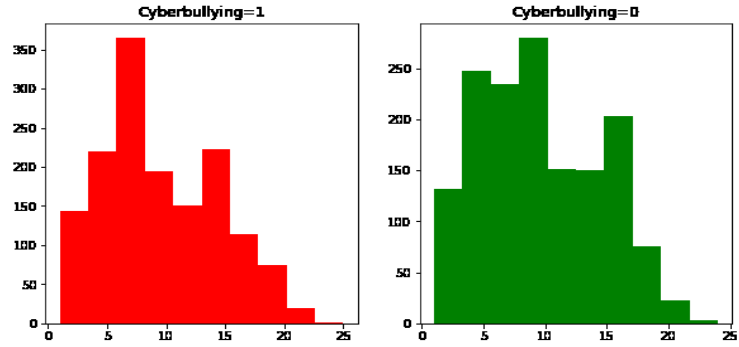

**Fig. 4.** Words in a tweet

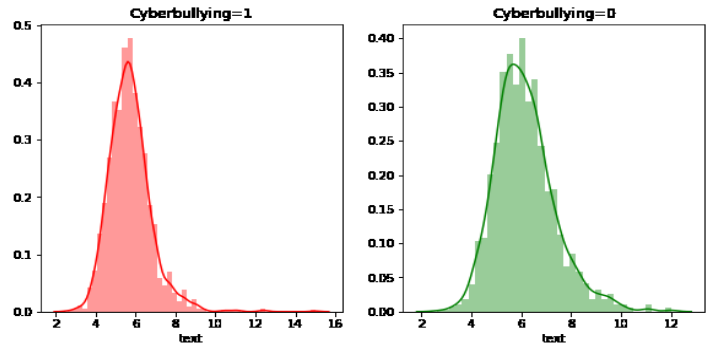

**Fig. 5.** Average word length of each tweet

In addition, among the tweets that contain cyberbullying and those that don't, an exploratory data analysis was performed in terms of characters, words and the average word length of each tweet, and the results obtained are shown in Fig.3, Fig.4 and Fig.5. In the experiment, it was observed that there is a formal similarity between the messages in Turkish that contain and do not contain cyberbullying. Moreover, the main limitation of the study is the use of machine learning algorithms for the specified data set.

**4.2. Data Preprocessing**

In data preprocessing, it is aimed to transform the texts in the data set into a common form. For these reasons, the tweets were converted to lowercase letters and the following unnecessary contents were removed from the text.

- Space Pattern
- URLs
- Punctuation
- Numbers and special characters
- Twitter Mentions

- Retweet Symbols
- Stopwords

After the unnecessary expressions in the text are cleaned out;

- Repeating Characters in Turkish words were determined and corrected (Correcting Repeating Characters) (For example, instead of "salaaaaaaaak", "salak" was written) (The English word for salak is "stupid").

- Libraries such as TextBlob, SymSpell, pyspellchecker and autocorrect can be used for English and many languages, while libraries such as Zemberek, ITU Turkish NLP Web Service API and turkishnlp can be used in Turkish (Alshemali and Kalita, 2019; Arıkan et al., 2019; Tekumalla and Banda, 2020). It is impossible to predict "better" with absolute reliability among the libraries in question, as the type of use or misspelling of the available language in the languages used will result in an almost infinite data set. There are many studies in which no library has produced perfect results due to the complex nature of language processing. It has been observed that stemming and lemmatization processes in Turkish texts are not as successful as the English language. One of the main reasons for this is that the prevalence of Turkish language is less than English, so there are no examples of successful word classification, and another reason is that Turkish is more difficult to analyse than English because it is a head final language. Moreover, since Turkish is a head final language, reduction to the root changes its meaning. For all these reasons, stemming and lemmatization have not been applied.

- Due to the widespread use of informal language in social media, the wrong or incomplete spellings of the words found to have bad meanings in their content were also determined, and 305 different spellings of 126 words were determined. For example, a system has been created in which Turkish words such as "qerizekali", "gerızekalı", "gerizekalı", "gerzekalı" are defined as "gerizekalı" ("gerizekalı" means "idiot" in English). In line with the determinations made, necessary corrections were made during the pre-treatment phase.

### 4.3. Feature Extraction

The most important point in classifying texts is to digitize the texts in accordance with machine learning models. In this study, the feature vector was created by applying the bag of words (BoW) method. In order to extract the BoW features, a vocabulary containing unigrams was created first and the terms with document frequency below 2 were ignored. Then, TF-IDF properties of the obtained unigrams were calculated. All operations are done using the Python Sklearn library.

### 4.4. Machine Learning Methods

Supervised machine learning models were applied to detect cyberbullying messages in social media messages written in Turkish. Any message containing cyberbullying is coded as 1 and not containing it as 0, a binary classification has been created. The data set is divided into 70% experimental (2099 Tweets) and 30% (901 Tweets) test data. Each tweet included in the training data set was tested on test data by trying to learn with machine learning models.

Nineteen (19) different machine learning algorithms (Gaussian Naive Bayes, Multinomial Naive Bayes, Bernoulli Naive Bayes, Decision Tree Model, Extra Trees Classifier, Linear Discriminant Analysis, Quadratic Discriminant Analysis, AdaBoost (Adaptive Boosting), Gradient Boosting Model, Random Forest, Logistic Regression Model, Perceptron Neural Network, Linear Support Vector Classification, XGBoost, K - Nearest Neighbours Model, Support Vector Machine, Stochastic Gradient Descent, Light Gradient Boosting Model and Voting Classifier Model) were used in classification for the detection of cyberbullying. In the Voting Classifier Model algorithm, the values of the other 18 different algorithms have been processed.

**4.5. Evaluation Measurements**

Since cyberbullying detection is a classification task, the value of accuracy generally comes to mind as the first evaluation scale. However, taking only the linear part in the classification is an intuitive approach, and if the accuracy is only one measurement, it is possible to obtain a maximum accuracy of 80%. In addition, the accuracy value can give very different results in unbalanced class distributions. This problem is solved by taking precision and recall measurements into consideration. With F1 Score, the most valid result is obtained by taking the harmonic averages of precision and recall.

**Table 1.** Confusion Matrix

| Confusion Matrix | | Actual Value | |
|---|---|---|---|
| | Status | Pozitives | Negatives |
| Predicted Value | Pozitives | | |
| | Negatives | | |

In the study, the success of learning was calculated over the confusion matrix as shown in Tab 1. From the obtained values, F1 Score, accuracy, Precision and recall values were found on the following equations.

$$Accuracy = \frac{TP + TN}{TP + FP + TN + FN} \qquad Precision = \frac{TP}{TP + FP}$$

$$Recall = \frac{TP}{TP + FN} \qquad F\ Score = \frac{2 * Precision * Recall}{Precision + Recall}$$

**4.6. Experimental Settings**

Python 3.8 was used in this study. All studies were carried out using Windows PC powered with 12 GB RAM. The algorithms used are programmed in Python, using Pycharm. Sklearn, NLTK, Numpy, Pandas and Matplotlib libraries were used in text processing and modelling. In determining the parameter values of the classification algorithms, hyper parameter optimization has been made for each algorithm.

**5. Result and Discussion**

Nineteen different machine learning algorithms were used in this study, which was conducted to determine cyberbullying in social media based on machine learning in the Turkish twitter sample. By trying to learn the data on the training data, it was estimated that the tweets included in the test data contained or did not contain cyberbullying.

**Table 2** Confusion Matrix values for Machine Learning Models

| Models | True Positive | False Negative | False Positive | True Negative |
|---|---|---|---|---|
| K - Nearest Neighbors Model (KNN) | 430 | 269 | 19 | 183 |
| Support Vector Machine (SVM) | 420 | 182 | 29 | 270 |
| Light Gradient Boosting Model (LGBM) | 417 | 51 | 32 | 401 |
| Stochastic Gradient Descent (SGD) | 416 | 85 | 33 | 367 |
| Random Forest | 415 | 74 | 34 | 378 |
| Voting Classifier Model | 412 | 59 | 37 | 393 |
| Decision Tree Model | 412 | 59 | 37 | 393 |
| XGBoost Classifier | 411 | 82 | 38 | 370 |
| Logistic Regression Model | 409 | 57 | 40 | 395 |
| Extra Trees Classifier | 406 | 46 | 43 | 406 |
| Linear Support Vector Classification | 405 | 55 | 44 | 397 |
| Gradient Boosting Model | 398 | 71 | 51 | 381 |
| Perceptron | 398 | 69 | 51 | 383 |
| Bernoulli Naive Bayes | 394 | 48 | 55 | 404 |
| AdaBoost Classifier | 390 | 65 | 59 | 387 |
| Multinomial Naive Bayes | 372 | 37 | 77 | 415 |
| Linear Discriminant Analysis | 321 | 118 | 128 | 334 |
| Quadratic Discriminant Analysis | 302 | 142 | 147 | 310 |
| Gaussian Naive Bayes | 257 | 68 | 45 | 231 |

Test data includes 451 cyberbullying tweets and 450 non-cyberbullying tweets. When the confusion matrix values are examined in Table 2 for evaluation;

- The modelling with the highest True Positive (TP) value is KNN,
- The algorithm with the highest False Negative (FN) value is KNN,
- QDA is the algorithm with the highest False Positive (FP)
- The modelling with the highest True Negative (TN) value is Multinomial Naive Bayes,
- As a result of a successful machine learning model, it has been observed that values of False Positive and False Negative are expected to be low, but the values in KNN, SVM, Linear Discriminant Analysis and Quadratic Discriminant Analysis are quite high.

**Table 3** Evaluation Results of Machine Learning Models

| Models | F1-score | Accuracy | Precision | Recall |
|---|---|---|---|---|
| Light Gradient Boosting Model (LGBM) | 90.949 | 90.788 | 90.856 | 90.795 |
| Extra Trees Classifier | 90.122 | 90.122 | 90.123 | 90.123 |
| Voting Classifier Model | 89.565 | 89.345 | 89.434 | 89.353 |
| Decision Tree Model | 89.565 | 89.345 | 89.434 | 89.353 |
| Logistic Regression Model | 89.399 | 89.234 | 89.286 | 89.24 |
| Linear Support Vector Classification | 89.109 | 89.012 | 88.033 | 88.016 |
| Random Forest | 88.486 | 88.013 | 88.307 | 88.028 |
| Bernoulli Naive Bayes | 88.44 | 88.568 | 88.579 | 88.566 |
| Stochastic Gradient Descent (SGD) | 87.579 | 86.903 | 87.392 | 86.923 |
| XGBoost Classifier | 87.261 | 86.681 | 87.027 | 86.698 |
| Perceptron | 86.9 | 86.681 | 86.737 | 86.688 |
| Multinomial Naive Bayes | 86.713 | 87.347 | 87.652 | 87.333 |
| Gradient Boosting Model | 86.71 | 86.459 | 86.528 | 86.467 |
| AdaBoost Classifier | 86.283 | 86.238 | 86.243 | 86.234 |
| Gaussian Naive Bayes | 81.978 | 81.198 | 81.386 | 81.178 |
| Support Vector Machine (SVM) | 79.924 | 76.582 | 80.034 | 76.638 |
| K - Nearest Neighbors Model (KNN) | 74.913 | 68.036 | 76.055 | 68.128 |
| Linear Discriminant Analysis | 72.297 | 72.697 | 72.708 | 72.693 |
| Quadratic Discriminant Analysis | 67.637 | 67.925 | 67.926 | 67.92 |

For the evaluation, F1 Score, accuracy, precision and recall values obtained from the confusion matrix data in table 2 are shown in table 3. As seen in Table 3, it was determined that the accuracy and F1 Score values of machine learning models in unsuccessful prediction were also low. The accuracy and F1 score values of the first 14 of 19 models are between 86% and 91%. In the modelling for detecting cyberbullying in Turkish tweets, the best result was determined using the LGBM algorithm with 90.949% F1 score and 90.788% accuracy. The worst classification result was obtained with the QDA algorithm with 67.637% F1 score and 67.925% accuracy.

Within the scope of the classification made for the detection of cyberbullying in Turkish texts;

- The performance of algorithms in tree-based and ensemble learning structure is good,
- The 90,949% F1 Score value obtained with the LGBM algorithm is the most successful result known in the Turkish literature using the classical machine learning model,
- It has been determined that the result obtained was similar to F1 Score obtained by Bozyiğit et al. (2019) in two-layer artificial neural networks,
- Especially, XGBoost and LGBM algorithms developed in recent years are the first known performance evaluation in Turkish cyberbullying texts, besides, they are not used frequently with texts in different languages and the results of the studies are similar to our research in terms of performance,
- When the present results are compared with the results obtained in the studies conducted in different languages in the relevant studies section, meaningful results could be obtained in Turkish as well.

## 6. Conclusion and Future Works

Social media gives its users the chance to express their feelings and thoughts on various topics. However, some users use social media maliciously with acts such as cyberbullying and express insults, threats and hate against other users. Within this context, machine learning models were used in our study to detect abuse in social media for malicious purposes. It is evaluated that it is different from many cyberbullying detection studies conducted in the international arena due to the fact that the cyberbullying detection study is conducted exclusively in Turkish language and also uses many up-to-date machine learning algorithms developed in recent years. Within the scope of the study, cyberbullying detection performances were tried to be determined by using 19 different machine learning algorithms over Turkish social media messages in the Twitter sample. The best result obtained by using the LGBM algorithm was 90.949% F1 score and 90.788% accuracy. The aforementioned result shows that cyberbullying based on machine learning can be successfully detected in Turkish, which has not been studied internationally.

Within this context, it is aimed to improve the data set regarding future studies, to include Turkish posts from Facebook and Instagram as well as Twitter and to increase the performance of the Turkish cyberbullying detection model by increasing the definition of bad words used in the Turkish natural language processing method.